KOENEMANN, F.H.

## Cauchy stress in mass distributions

*The thermodynamic definition of pressure $P = \partial U/\partial V$ is one form of the principle that in a given state, the mass in V and potential are proportional. Subject of this communication is the significance of this principle for the understanding of Cauchy stress.*

The stress theory as it is used today, was developed by Euler in 1776 and Cauchy in 1823. The following is a slightly edited quote of Truesdell [1].

    p.164: Let **f** be pairwise equilibrated; let -*S* denote the contact having the same underlying set as *S* but opposite orientation; then

$$\mathbf{t}_{-S} = -\mathbf{t}_S \tag{1}$$

    p.170: Cauchy assumed that the tractions **t** on all like-oriented contacts with a common plane at **x** are the same at **x**, i.e. $\mathbf{t}_S$ at **x** is assumed to depend on *S* only through the normal **n** of *S* at **x**: $\mathbf{t}_S = \mathbf{t}(\mathbf{x}, \mathbf{n})$. This statement is called the Cauchy postulate. *S* is oriented so that its normal **n** points out of c(*B*) if *S* is a part of $\partial$c(*B*). Thus $\mathbf{t}(\mathbf{x}, -\mathbf{n})$ is the traction at **x** on all surfaces **S** tangent to $\partial$c(*B*) and forming parts of the boundaries of bodies in the exterior c($B^e$) of c(*B*). In this sense $\mathbf{t}(\mathbf{x}, \mathbf{n})$ is the traction exerted upon *B* at **x** by the contiguous bodies outside it. As a trivial corollary of (1) follows Cauchy´s fundamental lemma: $\mathbf{t}(\mathbf{x}, -\mathbf{n}) = -\mathbf{t}(\mathbf{x}, \mathbf{n})$.

    p.176: $\mathbf{v}_1$ and $\mathbf{v}_2$ are linearly independent. At a given place $\mathbf{x}_0$ the planes $P_1$ and $P_2$ normal to $\mathbf{v}_1$ and $\mathbf{v}_2$, respectively, are distinct. We set $\mathbf{v}_3 = -(\mathbf{v}_1 + \mathbf{v}_2)$ and consider the wedge **A** that is bounded by these two planes and the plane $P_3$ normal to $\mathbf{v}_3$ at the place $\mathbf{x}_0 + \varepsilon\mathbf{v}_3$. We suppose $\varepsilon$ small enough that **A** be the shape of some part of *B*, and we denote by $\partial_i\mathbf{A}$ the portion of the plane $P_i$ that makes a part of the boundary of **A**. We let $\varepsilon$ approach 0. If we write $A_i$ for the area of $\partial_i\mathbf{A}$, we see that

$$A_1 = \frac{|\mathbf{v}_1|}{|\mathbf{v}_3|} A_3, \qquad A_2 = \frac{|\mathbf{v}_2|}{|\mathbf{v}_3|} A_3,$$
$$A_3 = O(\varepsilon) \quad \text{as} \quad \varepsilon \to 0, \tag{2}$$
$$V(\mathbf{A}) = \frac{\varepsilon|\mathbf{v}_3|A_3}{2}.$$

$$\text{If} \quad \mathbf{c} = \frac{|\mathbf{v}_3|}{A_3} \int_{\partial \mathbf{A}} \mathbf{t}(\mathbf{x}, \mathbf{n})\, dA, \tag{3}$$

from (2) and the assumption that $\mathbf{t}(\cdot, \mathbf{n})$ is continuous we see that

$$\mathbf{c} = \sum_{i=1}^{3} \frac{|\mathbf{v}_i|}{A_i} \int_{\partial_i \mathbf{A}} \mathbf{t}\!\left(\mathbf{x}, \frac{\mathbf{v}_i}{|\mathbf{v}_i|}\right) dA + O(\varepsilon) \quad \text{as} \quad \varepsilon \to 0. \tag{4}$$

Since **t** is a homogeneous function of its second argument and a continuous function of its first argument,

$$\mathbf{c} \to \sum_{k=1}^{3} \mathbf{t}(\mathbf{x}_0, \mathbf{v}_k) \quad \text{as} \quad \varepsilon \to 0. \tag{5}$$

On the other hand, we see that $\mathbf{c} \to \mathbf{0}$ as $\varepsilon \to 0$. Therefore, since the sum in (5) is independent of $\varepsilon$, it must vanish:

$$\sum_{k=1}^{3} \mathbf{t}(\mathbf{x}_0, \mathbf{v}_k) = \mathbf{0}. \text{ [End of quote]} \tag{6}$$

    The key argument in the above text is: "*since the sum in (5) is independent of $\varepsilon$*". It is an *a priori* condition; behind this is the assumption that Newton's 3rd law (1) is the proper equilibrium condition for the problem, and the Newtonian understanding of pressure, $P = |\mathbf{f}|/A$ which is believed to be universally scale-independent. Pressure is a state function, and a pressure increase requires that work is done on a system of mass distributed in *V*; it is a change of state in the sense of the First Law. Pressure is defined as energy density,

$$P = \partial U/\partial V. \tag{7}$$

The question is then: how are $P = |\mathbf{f}|/A$ and $P = \partial U/\partial V$ mathematically related?

The thermodynamic definition of $P$ is scale-independent, and an explicit statement of the proportionality of mass (measured in $V$ the radius of which is $r = |\mathbf{r}|$) and potential $U$ in a given state (Kellogg [2:80]). The thermodynamic equilibrium condition is

$$P_{surr} - P_{syst} = 0 \tag{8}$$

in scalar form. If both terms are thought to be caused by forces $\mathbf{f}$ [Newton] acting from either side on the surface of the system $V$ the equilibrium condition is

$$\mathbf{f}_{surr} - \mathbf{f}_{syst} = 0; \tag{9}$$

for isotropic conditions (subsequently implied), both $\mathbf{f}$ are radial force fields. Since the system contains mass, and since it interacts with the surrounding through exchange of work, it acts as a source of forces; i.e. its source density $\varphi \neq 0$ in some statically loaded state. $\varphi$ is always proportional to the mass in the system (Kellogg [2:45]); an existence theorem requires that if there is some function $f$ of a point $Q$ such that

$$\int f(Q)\,dV = \varphi, \tag{10}$$

both LHS and RHS must vanish with the maximum chord of $V$ (Kellogg [2:147]). As with all of thermodynamics (Born [3]), the approach to stress must thus be based on a Poisson equation (Kellogg [2:156]). The equilibrium condition (8) thus can take the form

$$\varphi_{surr} - \varphi_{syst} = 0. \tag{11}$$

It is therefore of interest how the volume functions relate to the surface functions if the domain of interest $V$ is changed in scale. In

$$\int \mathbf{f} \cdot \mathbf{n}\, dA = \int \nabla \cdot \mathbf{f}\, dV = \varphi, \tag{12}$$

$\mathbf{f}$ may be either one of the LHS terms in (9). If mass is continuously distributed, $\varphi \propto V$, and $\nabla \cdot \mathbf{f}$ is a constant that is characteristic of the energetic state in which the system is. Hence in (12), LHS $\propto V$. Since $V \propto r^3$, but $A \propto r^2$, for LHS $\propto r^3$ to hold it follows that $|\mathbf{f}| \propto r$, or

$$\frac{|\mathbf{f}|}{|\mathbf{r}|} = const. \tag{13}$$

Thus as $V \to 0$, $|\mathbf{f}|/A \to \infty$, yet $\Delta U/\Delta V \to const$. Both $\mathbf{f}_{syst}$ and $\mathbf{f}_{surr}$ vanish with $\mathbf{r}$; the condition in (10) is observed, stating that a system $V$ with zero magnitude cannot do work on its surrounding, and vice versa.

$\varepsilon$ (2-5) is an one-dimensional measure of the magnitude of the prism $\mathbf{A}$ (2), as is $r$ for $V$ in the subsequent discussion. It is to be taken into account that $P = |\mathbf{f}|/A$ is scale-independent if $A$ is a free plane, yet both the surface of the prism $\mathbf{A}$ in (2) and the surface $A$ in (12) are closed surfaces. The difference between (1) and (9) is that the latter distinguishes system and surrounding whereas the former does not. The thermodynamic system $V$ represents a distributed source in the sense of potential theory (Kellogg [2:150 ff]). $\varepsilon$ or $r$, respectively, is the zero potential distance (Kellogg [2:63]) which may have infinite length, or if it is finite it is set to have unit length by convention, but it cannot be zero or otherwise be let vanish.

*Address:* FALK H. KOENEMANN, Im Johannistal 36, 52064 Aachen, Germany; peregrine@t–online.de